%% file: Danilov_EPS-HEP2013.tex
\documentclass{PoS}

\usepackage{graphicx}

\title{Sensitivity of the DANSS detector to short range neutrino oscillations}

\ShortTitle{Sensitivity of the DANSS detector to neutrino oscillations}

\author{\speaker{Mikhail Danilov~$^{a,b}$}\\
  Representing the DANSS Collaboration (ITEP(Moscow) and JINR(Dubna))\\
  $^a$ ITEP -- State Scientific Center, Institute for
  Theoretical and Experimental Physics, Moscow, Russia\\
  $^b$ MIPT -- Moscow Institute of Physics and Technology, Moscow Region, Dolgoprudny, Russia\\
  E-mail: \email{danilov@itep.ru}}

\abstract{ DANSS is a highly segmented $1\,m^3$ plastic scintillator
  detector. It's 2500 scintillator strips have a Gd loaded reflective
  cover. Light is collected with 3 wave length shifting fibers per
  strip and read out with 50 PMTs and 2500 SiPMs.  The DANSS will be
  installed under the industrial 3~GW$_{\rm th}$ reactor of the
  Kalinin Nuclear Power Plant at distances varying from $9.7\,m$ to
  $12.2\,m$ from the reactor core. Tests of the detector prototype
  DANSSino demonstrated that in spite of a small size
  ($20\times20\times100$~cm$^3$) it is quite sensitive to reactor
  antineutrinos, detecting about 70 Inverse Beta Decay events per day
  with the signal-to-background ratio of about unity. The prototype
  tests have demonstrated feasibility to reach the design performance
  of the DANSS detector. The DANSS experiment will detect about 10
  thousand $\widetilde{\nu_e}$ events per day with a background below
  $\sim$1\%. Detector will be calibrated every day and its position will
  be changed frequently to reduce systematic errors. These features
  will provide a high sensitivity to reactor antineutrino oscillations
  to sterile neutrinos, suggested recently to explain a so-called
  "reactor anomaly". Data taking will start already next year.}


\FullConference{The European Physical Society Conference on High Energy Physics \\
  18-24 July, 2013\\
  Stockholm, Sweden}

\begin{document}

\section{Introduction}
Recent calculations \cite{Mueller, Huber} predict about 5\% higher
$\widetilde{\nu_e}$ flux from reactors than previous estimates. The
new smaller value of the neutron life time \cite{Serebrov} increases
the expected number of detected antineutrino by about 1\%.  The
discrepancy of new predictions with the measured antineutrino flux at
short distances is called a "reactor anomaly". It can be interpreted
as the signature of additional sterile neutrino states with mass
splittings of the order of $1eV^2$ \cite{Menton}. Moreover the
deficits in observed events from high-intensity neutrino sources used
to calibrate solar neutrino detectors can be also explained by this
hypothesis \cite{Giunti}. The observation of apparent $\nu_e$ and
$\widetilde{\nu_e}$ appearance in accelerator experiments
\cite{Mueller01, Aguilar} indicate similar mass-squared
splittings. However it is not clear whether the appearance and
disappearance experiments can be simultaneously explained by sterile
neutrinos \cite{Maltoni}. In this paper we discuss the sensitivity of
the DANSS detector \cite{DANSS} to short range reactor
$\widetilde{\nu_e}$ oscillations.

\section{Detector DANSS}
The DANSS collaboration plans to build a relatively compact neutrino spectrometer which does not contain any flammable or other dangerous liquids and may therefore be located very close to a core of a 3~GW$_{\rm th}$  industrial power reactor. Due to a high $\widetilde{\nu_e}$  flux ($\sim5\times10^{13}\; \bar\nu_e /{\rm cm}^2/{\rm s}$ at a distance of 11 m) it could be used for the reactor monitoring and neutrino oscillation studies.

The DANSS detector will consist of highly segmented plastic scintillator with a total volume of 1 m$^3$, surrounded with a composite shield of copper (Cu), lead (Pb) and borated polyethylene (CHB), and vetoed against cosmic muons with a number of external scintillator plates.

\begin{figure}[th]
 \input{DANSSino_Fig_Strips_Modules.tex}
 \caption{\footnotesize The basic element (left) and two of fifty intersecting modules (right) of the DANSS detector.}
 \label{Fig.DANSS_Modules}
\end{figure}
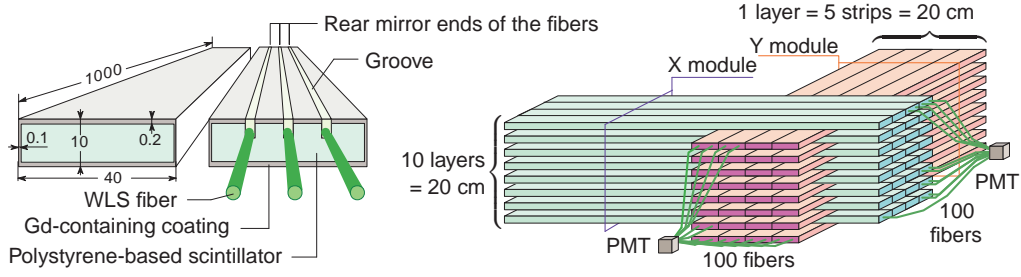

The basic element of DANSS is a polystyrene-based extruded scintillator strip ($1\times4\times100$~ cm$^3$) with a thin Gd-containing surface coating which is a light reflector and a ($n,\gamma$)-converter simultaneously (Fig.~\ref{Fig.DANSS_Modules}).
Light collection from the strip is done via three wavelength-shifting (WLS) Kuraray fibers Y-11, $\oslash$~1.2~mm, glued into grooves along the strip. One (blind) end of each fiber is polished and covered with a mirror paint, which decreases a total lengthwise attenuation of a light signal down to $\sim$20~\%/m.

Each 50 parallel strips are combined into a module, so that the whole detector (2500 strips) is a structure of 50 intercrossing modules (Fig.~\ref{Fig.DANSS_Modules}). Each module is viewed by a compact photomultiplier tube coupled to all 50 strips of the module via 100 WLS fibers, two per strip. In addition, to get more precise energy and space pattern of an event, each strip is equipped with an individual  SiPM (CPTA-151) coupled to the strip via the third WLS fiber. They provide about 10 photo-electrons (p.e.) per MeV in case of the DANSS strips. This number was obtained using measurements with cosmic rays. PMTs sensitivity determined from the calibration source measurements is also about 10 p.e./MeV.  

\section{The DANSSino design and tests}

In order to check operability of the DANSS design, to measure the real background conditions and shielding efficiency, a simplified pilot version of the detector (DANSSino) was constructed. 

\begin{figure}[ht]\label{Fig.DANSSino}
 \input{DANSSino_Fig_Construction.tex}
 \caption{\footnotesize The DANSSino detector}
 \end{figure}
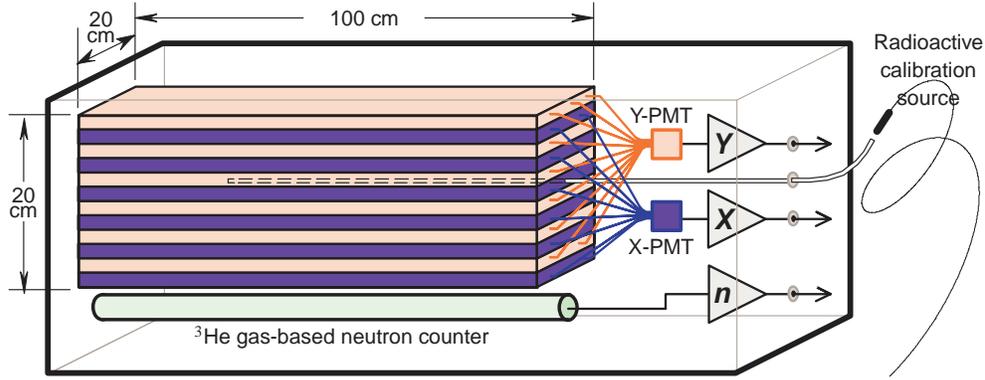

DANSSino consists of exactly the same basic elements as the main DANSS detector.
One hundred strips of DANSSino (Fig.~\ref{Fig.DANSSino}) form a bar $20\times20\times100$ cm$^3$ divided into two modules: the odd strip layers are coupled to the X-PMT and the even ones to the Y-PMT. Information from the individual SiPMs is not used in this prototype.
Energy calibration is performed using $^{137}$Cs, $^{60}$Co, $^{22}$Na, $^{248}$Cm sources. The detector is surrounded with a passive shield. A set of big scintillator plates ($200\times50\times3$~cm$^3$) form an active cosmic veto system. 

The Inverse Beta-Decay (IBD) of hydrogen in the detector body is used to detect $\widetilde{\nu_e}$.
\begin{equation}
 \widetilde{\nu_e}+p\rightarrow e^++n\;.
\end{equation}
The prompt signal is produced by the positron and two photons from its annihilation.  The neutron after moderation in the plastic scintillator is captured by $^{157}$Gd or $^{155}$Gd. In both cases a cascade of $\gamma$-rays is emitted with the total energy of about 8~MeV. The trigger is produced by any Prompt signal, and then the system waits for the Delayed signal during some fixed time. The energy of both Prompt and Delayed signals detected by both X and Y modules are measured. Thus each collected event contains 4 energies ($E_{XP}$, $E_{YP}$, $E_{XD}$, $E_{YD}$), time between the P and D pulses ($T_{PD}$) and information about the muon veto.

The DANSSino detector was installed at the Kalinin Nuclear Power Plant (350 km NW of Moscow) at a distance of 11 m from the WWER-1000 reactor core center. Materials above the detector provide excellent shielding ($\simeq$50 m w. e.) which completely removes fast cosmic neutrons. The muon component is suppressed by a factor of $\simeq$6. 
The results of some background measurements performed at the JINR laboratory and under the operating KNPP reactor with and without shielding are presented\footnote{The symbols $\wedge$ ({\sf AND}) and $\oplus$ ({\sf XOR}) here and below stand for logical operations of conjunction and exclusive disjunction, respectively.} in Table~\ref{Tab.BG}. Although the initial background conditions under the reactor are worse, a relatively thin passive shielding improves it significantly and makes even 3 times better than in the laboratory. The columns 5 and 6 reflect mainly the flux of thermal neutrons and cosmic muons, respectively (the neutrons produce X$\wedge$Y-coincidences with the total energy between the threshold and 8 MeV, whereas the muons also cause X$\wedge$Y coincidences but saturate the QDCs).

\begin{table}[ht]\centering
\caption{\footnotesize Background in the JINR laboratory and under the KNPP reactor measured by DANSSino unshielded and shielded with 10 cm of lead, 16 cm of CHB and $\mu$-veto plates (here and below the shielding composition is enumerated from inside to outside).}
\label{Tab.BG}
\input{DANSSino_Tab1.tex}
\end{table}

The last two columns of Table~\ref{Tab.BG} show the rate of events consisting of (Prompt+Delayed) signal pairs without any additional selection. A big number of these events for the unshielded detector arises from random coincidences caused by the high raw count rate. In the case of the shielded detector those are signals mainly from the fast neutrons. As expected, the number of these false neutrino-like events under the reactor becomes orders of magnitude lower than at the laboratory.
The remaining part is mostly associated with the $\mu$-veto signals and corresponds probably to secondary fast neutrons produced by cosmic muons in the surrounding heavy materials. Further background rejection is achieved by requirements motivated by the IBD signature:\\
1) the time between the Prompt and Delayed signals must be within a range $T_{PD}\in \left[1.5-30.0\right]\,\mu$s;\\
2) the Delayed signal should correspond to the Gd($n$,$\gamma$) reaction, i.e., both the X and Y modules should be fired ($X_D\wedge Y_D$) with a reasonable\footnote{As the detector is small, significant part of the $\gamma$-cascade is not detected, and therefore the acceptable $E_D$ range is extended to the lower energy.} total energy $E_{XD}+E_{YD}=E_D \in\left[1-8\right]$~MeV;\\
3) the Prompt energy must also be in a right range $E_P \in\left[1-7\right]$~MeV.

\begin{figure}[thb]
 \input{DANSSino_Fig_Run035_044.tex}
 \caption{\footnotesize Time dependence of the reactor power (bottom of each diagram) and the number of the neutrino-like events detected by DANSSino (top of each diagram) for two measurement periods.}\label{Fig.Nu(time)}
\end{figure}
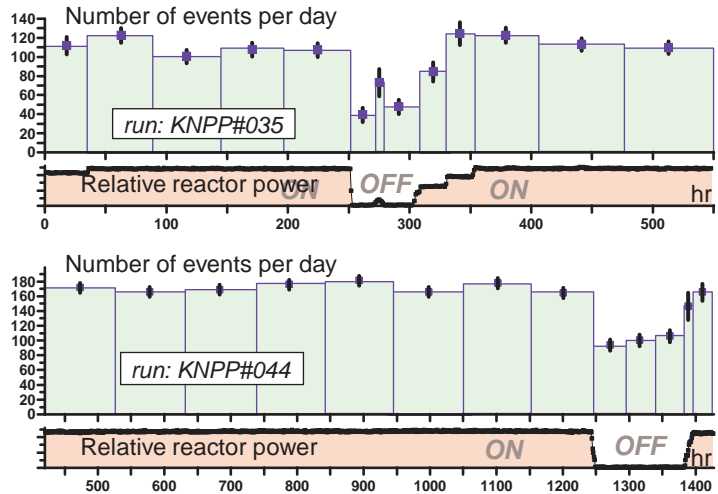

Two of our measurement runs came across few-days interruptions in the reactor operation (OFF periods), so that it was possible to estimate background experimentally without the neutrino flux. Fig.~\ref{Fig.Nu(time)} shows time diagrams of these runs. Strong correlation between the reactor power and the number of the neutrino-like events detected by DANSSino is obvious.
Numerical values of these count rates, Signal-to-Background ratios, and muon-induced background rates are given in Table~\ref{Tab.Nu(ON-OFF)}, which describes the neutrino-like events detected with an energy threshold $E_p^{\rm min}$= 1.5~MeV. It can be seen from the table that intensities of the muon-induced background events do not differ significantly for the ON and OFF periods, which indicates a low level of random coincidences. A heavy shielding without an inner moderator increases the background by $\sim$70\% but also improves the efficiency by $\sim$10\%, returning part of escaped IBD neutrons back to the scintillator.

\begin{table}[bht]\centering
\caption{Rate of the neutrino-like events in the reactor ON and OFF periods. The shielding thickness (cm) is given in parentheses.}
\label{Tab.Nu(ON-OFF)}
\begin{tabular}{|l||r|r||r|r|} \hline
Run\# & \multicolumn{2}{|c||}{ KNPP\#035 } & \multicolumn{2}{|c|}{ KNPP\#044 }\\ \hline
Shield composition & \multicolumn{2}{|c||}{\footnotesize\sf CHB(8)+Pb(10)}&
                \multicolumn{2}{|c|}{\footnotesize\sf Cu(5)+CHB(8)+Pb(5)} \\ \hline
Reactor operation &\multicolumn{1}{|c|}{OFF} &\multicolumn{1}{|c||}{ON} &\multicolumn{1}{|c|}{OFF}&\multicolumn{1}{|c|}{ON}\\ \hline \hline
Type of $\nu$-like events &\multicolumn{4}{|c|}{Number of $\nu$-like events per day}\\ \hline
Associated with $\mu$                 & 175$\pm$9&179$\pm$3&318$\pm$8&302$\pm$3\\ \hline
Free of $\mu$-veto                    &  46$\pm$5&108$\pm$3& 94$\pm$4&163$\pm$2\\ \hline \hline
Signal=ON--OFF & \multicolumn{2}{|c||}{62$\pm$5}&\multicolumn{2}{|c|}{70$\pm$5} \\ \hline              S/B=S/OFF      & \multicolumn{2}{|c||}{1.33$\pm$0.25}&\multicolumn{2}{|c|}{0.74$\pm$0.08} \\  \hline
\end{tabular}
\end{table}

\begin{figure}[tbh]
 \setlength{\unitlength}{1mm}
 \begin{picture}(150,52)(0,0)
 %\put(0,0){\framebox(150,52)[b]{}}
 \put(10,0){\includegraphics{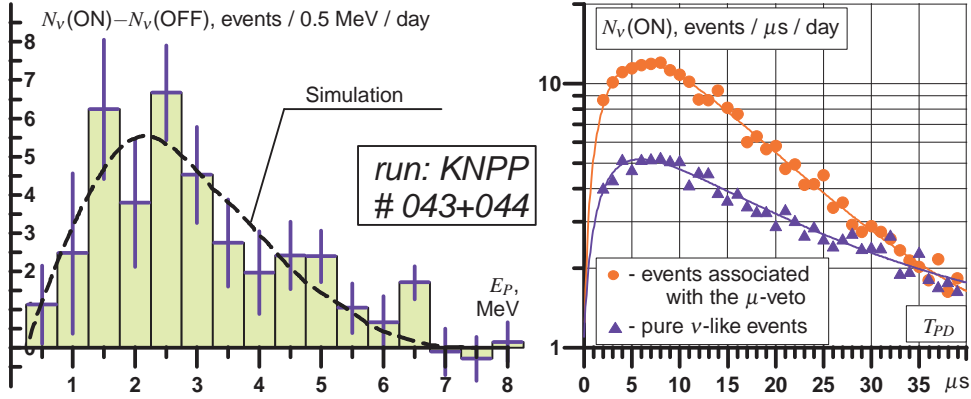}}
 \put(15.0,49.9){\makebox(0,0)[l]{\scriptsize\sf $N_\nu$(ON)$-N_\nu$(OFF), events / 0.5~MeV / day}}
 \put(78.5,14.5){\makebox(0,0)[r]{\scriptsize\sf $E_P$,}}
 \put(78.5,11.5){\makebox(0,0)[r]{\scriptsize\sf MeV}}
 \put(89.5,48.8){\makebox(0,0)[l]{\scriptsize\sf $N_\nu$(ON), events / $\mu$s / day}}
 \put(136.5,9.0){\makebox(0,0)[r]{\scriptsize\sf $T_{PD}$}}
 \put(138.3,0.3){\makebox(0,0)[rb]{\scriptsize\sf $\mu$s}}
 \put(50.0,39.0){\makebox(0,0)[lb]{\scriptsize\sf Simulation}}
 \put(93.0,16.2){\makebox(0,0)[l]{\scriptsize\sf- events associated}}
 \put(98.0,12.9){\makebox(0,0)[l]{\scriptsize\sf with the $\mu$-veto}}
 \put(93.0,09.0){\makebox(0,0)[l]{\scriptsize\sf- pure $\nu$-like events}}
 \end{picture}
 \caption{The differential $E_P$ energy spectrum (left) and $T_{PD}$
   time distribution (right) of the neutrino-like events. The dashed
   curve represents the IBD positron energy spectrum calculated for
   the $^{235}$U fission.}\label{Fig.Ep_Nu(ON-OFF)}
 \end{figure}

Assuming that the OFF data correspond to a pure background, one can build an energy spectrum of IBD positrons as a difference of two: $S$=$N_\nu$(ON)--$N_\nu$(OFF). As an example, $E_P$ and $T_{PD}$ spectra built under an additional ($X_P\!\wedge\! Y_P$) requirement\footnote{Otherwise, the low-energy part of the spectrum becomes polluted with random ($\gamma$-$n$) coincidences.} are shown in Fig.~\ref{Fig.Ep_Nu(ON-OFF)}. 
In spite of poor statistics and merely illustrative character of the spectra, they are in a very good agreement with theoretical expectations\footnote{For example MC simulations with GEANT-4 predict 75 events/day for the run \#044 while 70$\pm$5 events were observed.}. The delay time distribution for signal events agrees nicely with the expectations for the neutron capture process while for random coincidences the distribution should be flat.
The steeper slope of muon-induced events  is explained by the higher neutron multiplicity ($k\simeq1.6$).

\section{DANSS sensitivity to short range neutrino oscillations}

The agreement between MC expectations and actual DANSSino data provides confidence in the MC simulations of the DANSS detector and reliability of the estimated sensitivity to neutrino oscillations. DANSS efficiency for IBD decays is estimated to be about 70\%. The antineutrino counting rate will be as high as 10 thousand events per day. At the same time the background level is expected to be about 1\% only. These features together with a possibility to change frequently the distance from the reactor core make the DANSS very sensitive to neutrino oscillations. 

At this stage we perform a simple simulation of the detector response with the positron energy smeared only by the Poisson fluctuations in number of p.e. Errors in the energy measurement due to the light attenuation along the strip are smaller and in many cases can be corrected using information from the orthogonal strips. Non-uniformity across the strip is small enough to be neglected. For the sensitivity calculations we use so far only shape information of the measured spectra at different distances from the reactor core.

\begin{figure}[tbhp]
\includegraphics[width=0.45\linewidth]{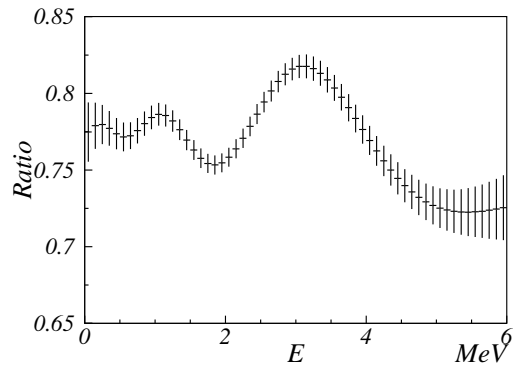}\centering
\caption{Ratio of positron energy spectra at 11~m and 9.7~m for $\Delta M^2=2eV^2$ and $sin^2(2\theta_{14})=0.2$. }
\label{Fig.Ratio}
\end{figure}

\begin{figure}[tbhp]
\setlength{\unitlength}{1mm}
\begin{picture}(150,85)
\put(30,5){\includegraphics[width=0.6\linewidth]{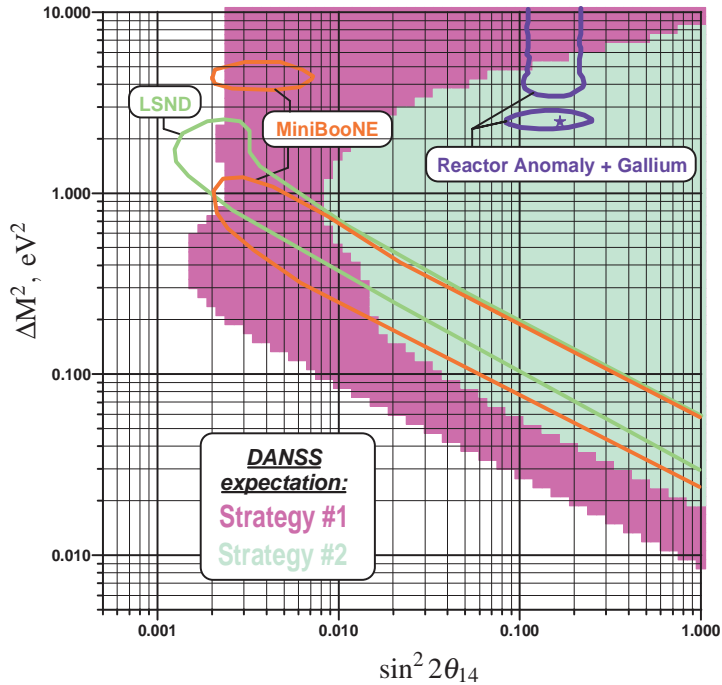}}
\put(75,0){$\sin^22\theta_{14}$}
\put(26,45){\rotatebox{90}{${\rm \Delta M^2,\;eV^2}$}}
\end{picture}
\caption{DANSS 95\% C.L. sensitivity contours for one year of running in case of the shape information only (green) and the most optimistic case of known neutrino energy spectrum (magenta).}\label{Fig.Sensitivity}
\end{figure}

The size of the core is quite big. However the oscillation pattern is smeared by this effect not too much. The ratio of positron energy spectra at 11 m and 9.7 m is shown in Fig.~\ref{Fig.Ratio} for the oscillation parameters $\Delta M^2=2eV^2$ and $sin^2(2\theta_{14})=0.2$. 
These parameters are close to the most probable values indicated by the "reactor anomaly". Errors in figure correspond to 8 months of running. The oscillation pattern is clearly seen. Using similar plots for other oscillation parameters the 95\% CL sensitivity region was obtained. It is shown in Fig.~\ref{Fig.Sensitivity}. 
It covers well the region indicated by the "reactor anomaly". So far
we have not included systematic errors in the estimates of the
sensitivity. However, they are expected to be very small because of
the very low background level, frequent changes of the detector
position, and its frequent calibration. We plan to start measurements
already in 2014. Thus the DANSS experiment has very good chances to
clarify the situation with the "reactor anomaly".

%\section{Acknowledgements}
The work was supported in part by the JINR grant 13-202-05, RFBR
grants 11-02-01251 and 11-02-12194, the RMES grants 8411, 1366.2012.2
and Rosatom contract H.4x.44.90.13.1119. We are grateful to the KNPP
Directorate and Radiation Safety Department for the constant support.

\end{document}

%% file: DANSSino_Fig_Strips_Modules.tex
 \setlength{\unitlength}{1mm}
 \begin{picture}(135,32)(0,3)
  %\put(0,0){\framebox(135,35)[b]{}}
  \put(0,1.0){\includegraphics{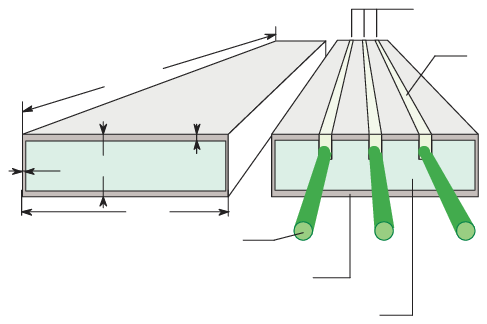}}
  \put(3.0,16.5){\makebox(0,0)[bl]{\tiny\sf 0.1}}
  \put(10.0,16.0){\makebox(0,0)[b]{\tiny\sf 10}}
  \put(14.5,12.0){\makebox(0,0)[c]{\tiny\sf 40}}
  \put(19.4,17.8){\makebox(0,0)[t]{\tiny\sf 0.2}}
  \put(11.3,25.1){\makebox(0,0)[c]{\tiny\sf 1}}
  \put(12.5,25.5){\makebox(0,0)[c]{\tiny\sf 0}}
  \put(13.8,25.9){\makebox(0,0)[c]{\tiny\sf 0}}
  \put(15.1,26.3){\makebox(0,0)[c]{\tiny\sf 0}}
  \put(23.0,8.0){\makebox(0,0)[br]{\scriptsize\sf WLS fiber}}
  \put(30.0,4.0){\makebox(0,0)[br]{\scriptsize\sf Gd-containing coating}}
  \put(37.0,0.0){\makebox(0,0)[br]{\scriptsize\sf Polystyrene-based scintillator}}
  \put(48.0,26.5){\makebox(0,0)[bl]{\scriptsize\sf Groove}}
  \put(42.5,32.5){\makebox(0,0)[l]{\scriptsize\sf Rear mirror ends of the fibers}}

  \put(66,1.5){\includegraphics{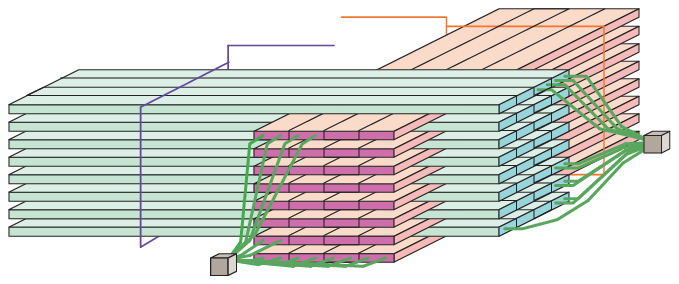}}
  \put(94.2,27){\makebox(0,0)[c]{\scriptsize\sf X module}}
  \put(105.0,29.6){\makebox(0,0)[c]{\scriptsize\sf Y module}}
  \put(67.0,12.5){\makebox(0,0.5)[r]{\scriptsize\sf $\left\{\rule{0mm}{8.2mm}\right.$}}
  \put(52.0,9.5){\parbox{12mm}
     {\begin{flushright}\scriptsize\sf 10 layers\\ = 20 cm\end{flushright}}}
  \put(129.0,32.5){\makebox(0,0)[rb]{\scriptsize\sf 1 layer = 5 strips = 20 cm}}
  \put(123.0,29.5){\makebox(0,0)[b]{\scriptsize\sf $\overbrace{\rule{15mm}{0mm}}$}}
  \put(83.0,2.0){\makebox(0,0)[b]{\scriptsize\sf PMT}}
  \put(99.0,0){\makebox(0,0)[b]{\scriptsize\sf 100 fibers}}
  \put(135,11.5){\makebox(0,0)[r]{\scriptsize\sf PMT}}
  \put(129.0,7.5){\makebox(0,0)[r]{\scriptsize\sf 100}}
  \put(130.0,4.5){\makebox(0,0)[r]{\scriptsize\sf fibers}}
 \end{picture}

%% file: DANSSino_Fig_Construction.tex
 \setlength{\unitlength}{1mm}
 \begin{picture}(135,45)(0,5)
  %\put(0,0){\framebox(135,50)[b]{}}
  \put(2,0){\includegraphics{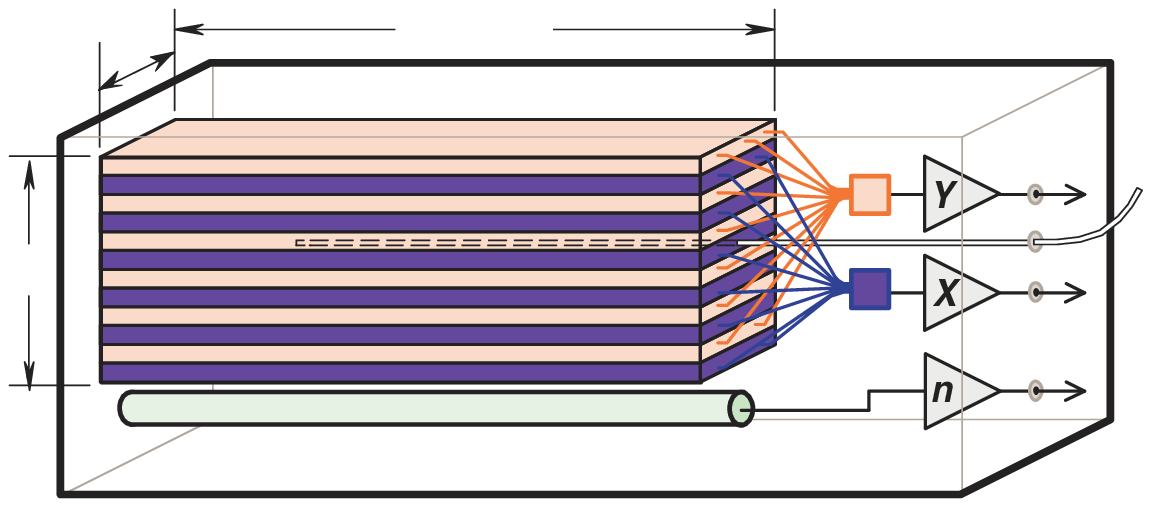}}
  \put(115.0,38.0){\parbox{20mm}{\begin{center}\scriptsize\sf Radioactive\\ calibration\\  source\end{center}}}
  \put(50.0,48.0){\makebox(0,0)[c]{\scriptsize\sf 100 cm}}
  \put(4.7,24.4){\makebox(0,0)[c]{\scriptsize\sf 20}}
  \put(4.7,22.2){\makebox(0,0)[c]{\scriptsize\sf cm}}
  \put(15.0,47.9){\makebox(0,0)[c]{\scriptsize\sf 20}}
  \put(15.0,45.5){\makebox(0,0)[c]{\scriptsize\sf cm}}
  \put(89.5,35.0){\makebox(0,0)[c]{\scriptsize\sf Y-PMT}}
  \put(89.5,17.5){\makebox(0,0)[c]{\scriptsize\sf X-PMT}}
 \put(47.0,4.0){\makebox(0,0)[b]{\scriptsize\sf $^3$He gas-based neutron counter}}
  \put(115.5,0){\includegraphics{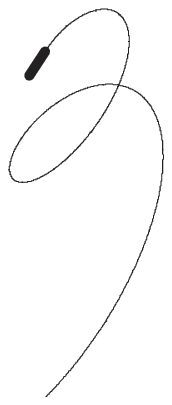}}
 \end{picture}

%% file: DANSSino_Tab1.tex
{\footnotesize
\begin{tabular}{|c|c||r|r||r|r||r|r|} \hline
Operation  & Detector \rule{0mm}{4mm}&\multicolumn{4}{|c||}{Module count rate,}
&\multicolumn{2}{|c|} {(P+D) pairs}\\
conditions & shielding \rule[-2mm]{0mm}{3mm}&\multicolumn{4}{|c||}{counts per second}
&\multicolumn{2}{|c|} {per day}\\ \cline{3-8}
&&\multicolumn{1}{|c|}{\scriptsize\sf X}
&\multicolumn{1}{|c||}{\scriptsize\sf Y}
&\multicolumn{1}{|c|}{\scriptsize\sf X$\wedge$Y}
&\multicolumn{1}{|c||}{\scriptsize\sf X$\wedge$Y}&&\\
&&\multicolumn{1}{|c|}{\scriptsize\sf E$\geq$0.25}
&\multicolumn{1}{|c||}{\scriptsize\sf E$\geq$0.25}
&\multicolumn{1}{|c|}{\scriptsize\sf E$\geq$0.5}
&\multicolumn{1}{|c||}{\scriptsize\sf E$\geq$8.0}
&\multicolumn{1}{|c|}{{\scriptsize\sf NO }$\mu$}
&\multicolumn{1}{|c|}{{\scriptsize\sf AND }$\mu$}\\
&&\multicolumn{1}{|c|}{\scriptsize\sf$\gamma$+$n$+$\mu$}
&\multicolumn{1}{|c||}{\scriptsize\sf$\gamma$+$n$+$\mu$}
&\multicolumn{1}{|c|}{\scriptsize\sf$n$+$\mu$}
&\multicolumn{1}{|c||}{\scriptsize\sf$\sim\mu$}&&\\ \hline\hline
  JINR& no shielding  \rule[-1.5mm]{0mm}{5.0mm}& 532 & 465 & 235  & 19 & \multicolumn{2}{|r|}{601 400\hspace*{5mm}} \\ \cline{2-8}
{\sf natural BG}\rule[-1.5mm]{0mm}{5.0mm}      & {\scriptsize\sf Pb+CHB+$\mu$-veto} & 61 & 58 & 42 & 17 & 30 750 & 9 030 \\ \hline \hline
 KNPP & no shielding \rule[-1.5mm]{0mm}{5.0mm} & 1 470 & 1 360 & 408  & 4 & \multicolumn{2}{|r|}{11 837 500\hspace*{5mm}} \\ \cline{2-8}
 {\scriptsize\sf $5\!\times\!10^{13}\nu$/cm$^2\!$/s}\rule[-1.5mm]{0mm}{5.0mm}&{\scriptsize\sf Pb+CHB+$\mu$-veto}& 20 & 19 & 11 & 2 & 1 240 & 980 \\ \hline
\end{tabular}
}

%% file: DANSSino_Fig_Run035_044.tex
 \setlength{\unitlength}{1mm}
 \begin{picture}(135,58)(-30,5)
  %\put(0,0){\framebox(135,67)[b]{}}
%  \put(0,35){\includegraphics{Fig_R035_100.eps}}
%  \put(0,35){\includegraphics[scale=0.7]{Fig_5a.eps}}
  \put(0,37){\includegraphics[scale=0.7]{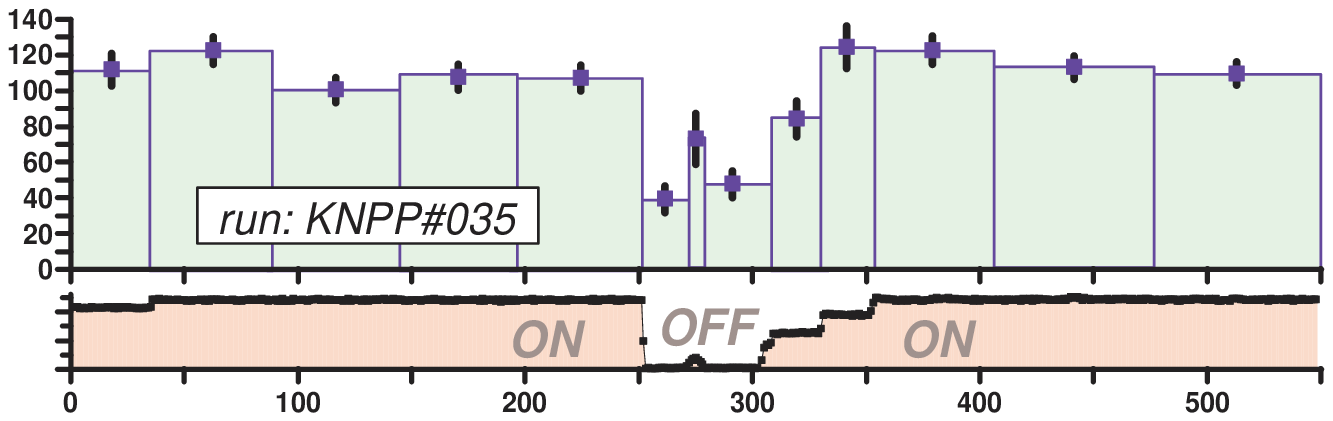}}
  \put( 7.5,66.9){\makebox(0,0)[lt]{\footnotesize\sf Number of events per day}}
  \put( 9.5,43.0){\makebox(0,0)[l]{\footnotesize\sf Relative reactor power}}
  \put(93.5,42.0){\makebox(0,0)[r]{\footnotesize\sf hr}}
%  \put(0,0){\includegraphics{Fig_R044_100.eps}}
%  \put(0,0){\includegraphics[scale=0.7]{Fig_5b.eps}}
  \put(0,2){\includegraphics[scale=0.7]{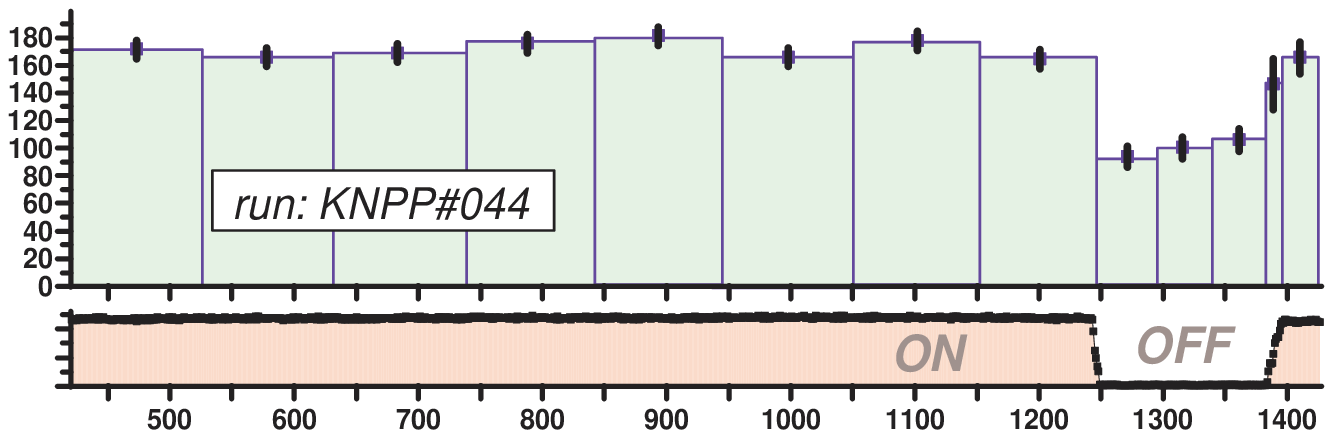}}
  \put( 7.5,33.9){\makebox(0,0)[lt]{\footnotesize\sf Number of events per day}}
  \put( 9.5,8.0){\makebox(0,0)[l]{\footnotesize\sf Relative reactor power}}
  \put(93.5,6.9){\makebox(0,0)[r]{\footnotesize\sf hr}}
 \end{picture}